\documentclass[journal=jctcce,manuscript=article,layout=twocolumn]{achemso}
\setkeys{acs}{articletitle=true}

\usepackage[T1]{fontenc}
\usepackage{amsmath, amssymb}
\usepackage{siunitx}
\usepackage{xcolor}
\usepackage{braket}
\usepackage{multirow}
\usepackage[version = 3]{mhchem}
\usepackage{paralist}
\usepackage[normalem]{ulem}

\title{Optimally Tuned Multiconfigurational Short-Range DFT for Linear Response Properties}
\author{Micha{\l} Hapka}
\email{michal.hapka@uw.edu.pl}
\affiliation{University of Warsaw, Faculty of Chemistry, ul.\ L.\ Pasteura 1, 02-093 Warsaw, Poland}
\author{Katarzyna Pernal}
\affiliation{Institute of Physics, Lodz University of Technology, ul. Wolczanska 217/221, 93-005 Lodz, Poland}
\author{Ewa Pastorczak}
\email{ewapastorczak@gmail.com}
\affiliation{Institute of Physics, Lodz University of Technology, ul. Wolczanska 217/221, 93-005 Lodz, Poland}

\begin{document}

\begin{abstract}
Multiconfigurational short-range density functional theory (MC-srDFT) rigorously combines ground state wavefunction theory with DFT. Unlike single-reference range-separated hybrid functionals, MC-srDFT has lacked theoretically-grounded protocols for choosing the system-specific range-separation parameter. To address this problem, we introduce an optimal-tuning scheme based on enforcing the correct exponential decay of the electron density. We~show that the range-separation parameter can be determined from the ionization potential given by the smallest-magnitude eigenvalue of the Extended Koopmans’ Theorem matrix constructed for the model Hamiltonian. We validate this approach for static and dynamic dipole polarizabilities of ground-state molecular systems using MC-srDFT within both full linear response and its extended random phase approximation (ERPA) variant. Optimal tuning substantially improves polarizabilities relative to the commonly used universal $\mu=0.4$~bohr$^{-1}$ parameter. 
\end{abstract}

\maketitle

\section{Introduction} \label{intro}

Wavefunction theory (WFT) and DFT can be rigorously combined in a way that benefits both approaches, even when approximate treatments are employed. A prominent example is provided by range-separated methods, in which the electron–electron interaction is decomposed into short- and long-range components treated by DFT and WFT, respectively.~\cite{Hedegard:20,pernal2022range} This leads to range-separated multiconfigurational DFT (MC-srDFT), originally proposed by Savin et al.\cite{Stoll:85,Savin:95,Savin:96} to address the inability of standard DFT approximations to describe ground states of near-degenerate systems. Soon after its introduction, MC-srDFT was recognized as an efficient route for incorporating dynamic correlation into wavefunction methods.~\cite{Savin:96} By assigning short-range correlation to density functionals, the required wavefunction expansions can be kept compact, thereby reducing sensitivity to the choice of active space in complete active space (CAS) methods and to excitation-level truncation in configuration interaction\cite{Leininger:97,Ferte:19} and coupled-cluster\cite{Goll:05,Goll:06} theories.

To date, most MC-srDFT applications have focused on ground-\cite{Fromager:10,Hedegard:15,Hedegaard:18} and excited-state energies,\cite{fromager2013multi,pernal2012excitation,Hubert:16,Hubert:16b,Hedegard:17,Kjellgren:19,hapka2020long} demonstrating promising accuracy. Nevertheless, a central challenge of the approach---the dependence on the range-separation (RS) parameter $\mu$ that governs the partitioning of the Coulomb operator---has not been fully resolved. Fromager et al.\cite{Fromager:07} advocated the use of a universal value of $\mu$, chosen to preserve single-reference character in systems dominated by dynamic correlation while recovering essential static correlation effects in the multireference regime. Based on studies of ground states of small atoms and molecules,~\cite{Fromager:07,Fromager:09} they recommended $\mu = 0.4\ \mathrm{bohr}^{-1}$. On the one hand, the use of a universal, global $\mu$ guaranties size consistency and avoids system-specific parameter tuning. On the other hand, a systematic protocol for such tuning has not been developed within MC-srDFT. This is in stark contrast to single-reference range-separated DFT, where optimal tuning strategies are well established. These include either ionization potential (IP) tuning,\cite{Baer:10,Autschbach:14} which enforces the Koopmans condition,\cite{levy1984exact,almbladh1985exact} $\varepsilon_{\text{HOMO}}^{\text{KS}}=-\text{IP}$, which is exact in Kohn-Sham (KS) DFT, or global density-dependent tuning\cite{Modrzejewski:13,Mandal:25} in which the range-separation parameter becomes a functional of the density, $\mu\equiv\mu[\rho]$.

Relatively few works so far have employed time-dependent linear-response extensions of MC-srDFT\cite{fromager2013multi,pernal2014accurate} to the calculation of response properties. Hedeg\r{a}rd assessed the accuracy of CAS-srDFT oscillator strengths\cite{Hedegard:17} for the Thiel set\cite{Thiel:08} obtaining satisfactory agreement with CC2 results. Jensen and co-workers\cite{Kjellgren:21,Hapka:25} showed that both CAS-srDFT and GVB-srDFT (GVB denoting generalized valence bond wavefunction) give reliable predictions of indirect spin–spin coupling constants. Their work is a good example of MC-srDFT sensitivity to the RS parameter value: for transition-metal complexes, satisfactory results were obtained by setting $\mu$ to 1.0~$\mathrm{bohr}^{-1}$ instead of the standard 0.4~$\mathrm{bohr}^{-1}$ choice.

In this work, we introduce a rigorous and physically motivated procedure for determining the range-separation parameter, which, to the best of our knowledge, has not been previously formulated within the MC-srDFT framework. The proposed tuning criterion is based on an exact extension of the Koopmans condition to MC-srDFT and on the requirement that optimally tuned MC-srDFT, employing approximate short-range exchange–correlation functionals and approximate wavefunctions, reproduces the correct long-range asymptotic decay of the electron density.

We assess the performance of the optimally-tuned MC-srDFT approach by computing static and dynamic molecular polarizabilities. Dipole polarizabilities are known to be sensitive to both the quality of the electronic structure description and the employed basis set.~\cite{hait2018accurate,wilkins2019accurate} In particular, experience with range-separated hybrid (RSH) DFT functionals demonstrates a pronounced dependence of polarizabilities on the value of the RS parameter,\cite{Sun:13,Nenon:14,Oviedo:16,Alipour:17} which becomes even more significant for hyperpolarizabilities.~\cite{Garrett:14,Garza:15,Wang:17,Zalesny:19,Lescos:20,Besalu-Sala:20} It is worth noting that asymptotically corrected exchange-correlation potentials\cite{Tozer:98,Gruning:01,Cencek:13} offer an alternative way of improving DFT response properties by enforcing the correct density decay. Such potentials have found particularly extensive application within symmetry-adapted perturbation theory.~\cite{Jansen:14,Patkowski:20}

In addition to dipole polarizabilities computed from the full linear-response MC-srDFT formalism,\cite{fromager2013multi} we examine the performance of the extended random phase approximation (ERPA)\cite{erpa1,pernal2014accurate,ac0d} applied to MC-srDFT wave functions. In contrast to the full linear response, ERPA neglects contributions from the response of the wave function expansion coefficients, accounting solely for orbital relaxation. This simplification leads to a more favorable scaling with respect to the size of the active space.

The paper is organized as follows. In Section~\ref{sec:theory}, we 
briefly recapitulate the MC-srDFT formalism and introduce the proposed optimal tuning protocol. We then present ERPA-based linear response equations for the calculation of dipole polarizabilities in the MC-srDFT framework. 
In Section~\ref{sec3}, the performance of the employed linear-response approaches is benchmarked against CC3 dipole polarizabilities reference data for the test set of J{\o}rgensen et al.~\cite{sauer_polari}. Finally, Section~\ref{sec:summary} summarizes the main findings and outlines perspectives for further methodological development and applications.

\section{Theory}\label{sec:theory}
\subsection{Range-separated multicofigurational DFT (MC-srDFT)}

The multiconfigurational density functional theory, denoted here as MC-srDFT, proposed by Savin et al.\cite{Stoll:85,Savin:96,Pollet:2002p2162,Toulouse:2004}, cf.\ also Refs. \citenum{Hedegard:20} and \citenum{pernal2022range}, is based on the separation of the electron-electron Coulomb interaction operator,
$r_{12}^{-1}$, into short-range (SR) and long-range (LR) components, $\hat{\upsilon}_{ee}^{\rm SR}(r_{12})$ and $\hat{\upsilon}_{ee}^{\rm LR}(r_{12})$, respectively.
The essential conditions for the long-range part are that it is finite at the
electron-electron coalescence, $r_{12}\rightarrow0$, and reduces to the Coulomb interaction, $1/r_{12}$, in the large-separation limit.
The exact ground-state range-separated density functional takes the form
\begin{equation}
E^{\text{MC-srDFT}}[\Psi] = \left\langle \Psi|\hat{T}+\hat{V}_{ne}+\hat{V}_{ee}^{\rm LR}|\Psi\right\rangle + E^{\rm SR}[\rho_{\Psi}]
\label{MCFun}
\end{equation}
where $\hat{T}$ and $\hat{V}_{ne}$ are, respectively, the kinetic energy and
electron-nuclei interaction operators, $\rho_{\Psi}$ stands for electronic
density corresponding to a wavefunction $\Psi$. The short-range functional $E^{\rm SR}[\rho]$ is rigorously defined\cite{Toulouse:2004} but its explicit exact form
is unknown. For practical approximations,  $E^{\rm SR}[\rho]$ is typically split into a Hartree functional, $E_{\rm H}^{\rm SR}[\rho]=\frac{1}{2} \int\int \rho(\mathbf{r}_{1}) \hat{\upsilon}_{ee}^{\rm SR}(r_{12}) \rho(\mathbf{r}_{2}) \, \mathrm{d}\mathbf{r}_{1} \mathrm{d}\mathbf{r}
_{2}$,
and the exchange-correlation functional $E_{xc}^{\rm SR}[\rho]$ \cite{Toulouse:2004,Toulouse:05}.

Ground state energy follows from minimization of the MC-srDFT\ functional, which
is equivalent to solving an eigenproblem for the Hamiltonian $\hat{H}^{\rm LR}$
\begin{equation}
\hat{H}^{\rm LR}\ \Psi^{\rm LR} = \mathcal{E}^{\rm LR}\ \Psi^{\rm LR}\ \ \ 
\label{LRSE}
\end{equation}
where $\hat{H}^{\rm LR}$ includes the long-range electron-electron interaction operator
\begin{equation}
\hat{H}^{\rm LR} = \hat{T} + \hat{V}_{ne} + \hat{V}_{ee}^{\rm LR} + \hat{V}^{\rm SR}[\rho
_{\Psi^{\rm LR}}]\ \ \ 
\label{LRH}
\end{equation}
and the short-range local potential $\hat{V}^{\rm SR}$, which follows as a functional derivative of $E^{\rm SR}[\rho]$.
The wavefunction $\Psi^{\rm LR}$ differs from the exact ground state correlated wavefunction (unlike the latter, for example, the former is free of the electron coalescence cusp). Contrary to this, the electron density $\rho_{\Psi^{\rm LR}}$ corresponding to the wavefunction $\Psi^{\rm LR}$ coincides with the exact ground state density, $\rho_{0}$, as in the conventional DFT 
\begin{equation}
\rho_{0}(\mathbf{r)} = \rho_{\Psi^{\rm LR}}(\mathbf{r)\ \ \ } \label{rho0}
\end{equation}
on the condition that the SR\ functional is exact.

The most successful approximations to the SR exchange-correlation
functionals have been developed for the  error-function representation of the LR electron interaction\cite{paziani2006local,goll2009development,Goll:06,Ferte:19}
\begin{equation}
\hat{\upsilon}_{ee}^{\rm LR}(r_{12})=\frac{\operatorname{erf}(\mu r_{12})}{r_{12}%
}\ \ \  \label{erf}%
\end{equation}
which involves a range-separation parameter, $\mu$. By varying $\mu$ from $0$ to $\infty$, one smoothly switches between the KS-DFT\ and full-range-interaction-FCI limiting cases.

In contrast to MC-srDFT with the exact SR functional, which yields the same total energy for any value of the range-separation parameter, the use of approximate functionals makes MC-srDFT predictions dependent on its choice. In~principle, more accurate functional approximations should be less sensitive to the choice of $\mu$. 
Early investigations of the energies of atoms, small molecules, and dissociation energy curves suggested the value of the ``universal'' $\mu$ equal to $0.4$~a.u.\cite{Fromager:07}. In practice, choosing the optimal value of $\mu$ has remained an open problem, especially for excited-state energies\cite{Hedegaard:18,hapka2020long} and for the response properties\cite{Kjellgren:21,Hapka:25} such as molecular polarizability.

\subsection{Optimal range-separation parameter for MC-srDFT based on an exact condition}

Recall that if the exact SR\ functional were available, the wavefunction $\Psi^{\rm LR}$ resulting from MC-srDFT, see Eq.~\eqref{LRSE}, would still be different from the FCI\ wavefunction corresponding to the full-range electron interaction. However, the ground state electron density would be exact, see Eq.~\eqref{rho0}. It would therefore decay exponentially as
\begin{equation}
\rho_{\Psi^{\rm LR}}(r\rightarrow\infty)\sim \exp \left[ -2\sqrt{2\,\smash{\text{IP}_{1}}}r \right]  \label{as}
\end{equation}
Importantly, the IP$_{1}$ denotes the first ionization potential of the real system, i.e.\ the one described by the Hamiltonian with full-range electron interaction. It is therefore different from the ionization potential of a reference system described with a model Hamiltonian, $\hat{H}^{\rm LR}$.

Using approximate SR functionals in MC-srDFT violates the equality assumed in Eq.~\eqref{rho0} and, in consequence, yields erroneous asymptotic decay of the density.
By using arguments of the seminal paper by Morrell, Parr, and Levy (MPL)\cite{mpl}, one can show how to tune the RS parameter $\mu$ so that the approximate electron density decays exponentially with the exact ionization potential of a given system. 

We begin by generalizing derivations presented by MPL to model Hamiltonians. Assume a model ($M$) Hamiltonian involving a modified electron interaction
operator $\upsilon_{ee}^{M}$
\begin{equation}
\hat{H}^{M} = \sum_{i=1} \left( \hat{t}(\mathbf{r}_{i}) + \hat{\upsilon}_{ext}^{M}(\mathbf{r}_{i})\right) + \frac{1}{2} \sum_{i\neq j} \upsilon_{ee}^{M} (\left \vert \mathbf{r}_{i}-\mathbf{r}_{j} \right\vert )
\end{equation}
giving rise to a model electron density
\begin{equation}
\rho^{M}(x)=N\int|\Psi^{M}(x,x_{2},...,x_{N})|^{2} \, \mathrm{d}x_{2}\ldots \mathrm{d}x_{N}
\end{equation}
where $\Psi^{M}$ is a wavefunction corresponding to $\hat{H}^{M}$, and $x$ denotes a combined spin-spatial coordinate.

Provided that ($a$) $\hat{\upsilon}_{ext}^{M}(\mathbf{r})$ is a local potential vanishing at $r\rightarrow\infty$, and ($b$) $\upsilon_{ee}^{M}(r)$ decays as a Coulomb potential in the long-range
\begin{equation}
\upsilon_{ee}^{M}(r\rightarrow\infty)\sim\frac{1}{r}
\end{equation}
one can repeat the arguments of MPL [see Eqs.(43)-(58) therein] and arrive at
the asymptotic decay of a model electron density reading
\begin{equation}
\rho^{M}(r\rightarrow\infty) \sim \exp \left[ -2\sqrt{-2\lambda_{\text{max}}^{M}} \, r \right]  \label{asM}
\end{equation}
On the same basis, the long-range bahavior of the natural orbitals (NOs) $\left\{ \varphi_p^M(x)\right\}$, diagonalizing a model one-electron reduced density matrix (1-RDM), $\gamma^{M}$
\begin{align}
\gamma^{M}(x,x^{\prime}) & = \sum_{p} n_p^M \varphi_p^M (x^{\prime})^{\ast} \varphi_p^M(x)
\end{align}
is obtained, reading
\begin{align}
\forall_{p}\ \ \ \varphi_{p}^{M}(r & \rightarrow\infty) \sim \exp \left[
-\sqrt{-2\lambda_{\text{max}}^{M}} \, r \right]
\end{align}
The number $\lambda_{\text{max}}^{M}$ is the largest (least negative)  eigenvalue of the matrix $\mathbf{\Lambda}^{M}$ defined for a given model Hamiltonian. In the natural-orbitals representation, it takes the form
\begin{align}
\sqrt{n_{p}^{M}n_{q}^{M}}\Lambda_{pq}^{M} &= n_{q}^{M}\left\langle \varphi_{p}^{M}|\hat{t}+\hat{\upsilon}_{ext}^{M}|\varphi_{q}^{M} \right\rangle \nonumber \\ 
& +2\sum_{rst}\Gamma_{stqr}^{M}\ \left\langle \varphi_{p}^{M}\varphi_{r}^{M}|\upsilon_{ee}^{M}|\varphi_{s}^{M}\varphi_{t}^{M}\right\rangle \label{Lam} \\ \nonumber
\lambda_{\text{max}}^{M} &= \max \, \{ \lambda_p \}
\end{align}
where $\lambda_p$ are eigenvalues of $\mathbf{\Lambda}^{M}$, $\left\{ n_{p}^{M}\right\}$ denotes natural occupation numbers, and $\Gamma^{M}$ is the two-particle reduced density matrix, all corresponding to the model wavefunction $\Psi^{M}$.
In Eq.~\eqref{Lam}, we adopt physicist notation for two-electron integrals.  The matrix $\mathbf{\Lambda}^{M}$ has been known as the Extended Koopmans' Theory (EKT)
matrix.\cite{morrison1992extended,smith1975extension} Thus, any model density corresponding to a Hamiltonian with modified electron interactions of the Coulomb tail decays exponentially, and
the pace of this decay is governed by the first EKT\ eigenvalue which, in general, is specific to the assumed model.

Let us apply the general considerations for model Hamiltonians to MC-srDFT. By combining Eqs.~\eqref{as} and \eqref{asM}, we obtain an exact and nontrivial result that the largest eigenvalue, $\lambda_{\text{max}}^{M}$, of the
LR-EKT\ matrix, i.e.\ the matrix given by Eq.~\eqref{Lam} computed with a model
Hamiltonian matrix and reference RDMs corresponding to a model wavefunction
$\Psi^{\rm LR}$, is equal to IP$_{1}$ of a Coulomb system
\begin{equation}
-\lambda_{\text{max}}^{\rm LR} = \text{IP}_{1} \label{cond0}
\end{equation}
This property can be seen as a multiconfigurational generalization of the HOMO condition, i.e.\ equality of the negative HOMO orbital energy in the Kohn-Sham model and the exact ionization potential,\cite{levy1984exact,almbladh1985exact}
$-\varepsilon_{\rm HOMO}^{\rm KS}=$IP$_{1}$. In fact, in the limit $\hat{\upsilon}_{ee}^{\rm LR}(r_{12})\rightarrow0$, MC-srDFT\ turns to KS-DFT, eigenvalues of LR-EKT are just energies of the occupied KS\ orbitals, in particular $\lambda_{\text{max}}^{\rm LR}=\varepsilon_{\rm HOMO}^{\rm KS}$, and the condition in Eq.~\eqref{cond0} is trivially satisfied. We have just shown, however, that it is satisfied for any form of $\hat{\upsilon}_{ee}^{\rm LR}(r_{12})$.

Another feature of MC-srDFT, not fully recognized although straightforward to show from the discussion above, concerns the natural orbitals, $\left\{\varphi_{p}^{\rm LR}\right\}$, that diagonalize the 1-RDM pertaining to $\Psi^{\rm LR}$. These orbitals are neither KS\ orbitals nor exact natural orbitals corresponding to the full-range electron interaction, except in the limit $\hat{\upsilon}_{ee}^{\rm LR}(r_{12})\rightarrow\infty$. Following MPL, one concludes that they decay asymptotically with $\lambda_{\text{max}}^{\rm LR}$. Taking Eq.~\eqref{cond0} into account, this leads to the following asymptotic behavior 
\begin{equation}
\varphi_{p}^{\rm LR}(r\rightarrow\infty) \sim \exp \left[ -\sqrt{-2\text{IP}_{1}} \, r \right]  \label{NO}
\end{equation}
 Consequently, all MC-srDFT natural orbitals possess the same rate of decay as the exact ones. 

The analysis above pertains to exact MC-srDFT. We now consider an MC-srDFT model employing an approximate SR functional with the erf-modified electron interaction, Eq.~\eqref{erf}.  Imposing the condition in Eq.~\eqref{cond0}, which is satisfied by exact MC-srDFT, uniquely fixes the range-separation parameter $\mu$. 
The optimally tuned parameter $\mu_{\rm opt}$ satisfies the equation
\begin{equation}
-\lambda_{\text{max}}^{\rm LR}(\mu_{\rm opt}) = \text{IP}_{1} 
\label{cond}
\end{equation}
Whether  IP tuning is feasible depends on the particular form of
$\hat{H}^{M}(\mu)$. If it leads to densities that differ significantly from the exact one, 
Eq.~\eqref{cond} may have no solution.

While the improved accuracy of computations employing the optimal value of $\mu$ instead of the conventional value of $0.4$ a.u. is not guaranteed, we show in the next section that this is indeed the case for ground-state polarizability calculations. The improvement is likely related to the fact that the natural orbitals obtained from $\Psi^{\rm LR}$ with $\mu_{\rm opt}$ show a proper LR decay, see Eq.~\eqref{NO}, resembling the exact NOs' LR behavior.

\subsection{Approximate dynamic linear response from MC-srDFT}

The ground-state MC-srDFT theory can be extended to the time-dependent (TD) regime. The exact TD-MC-srDFT formalism was derived by Fromager and co-authors\cite{fromager2013multi} in the framework of the Floquet theory. Implemented in the adiabatic approximation for the short-range exchange-correlation kernel,\cite{pernal2012excitation} it has given rise to time-dependent MC-srDFT linear response equations.

Another route to deriving an approximate linear-response TD-MC-srDFT function is to employ the equations-of-motion formalism of Rowe\cite{rowe} in the extended random phase approximation.~\cite{erpa1,pernal2014accurate} Since our focus is on second-order response properties, we present below the relevant equations in the ERPA-MC-srDFT framework.

Consider a frequency-dependent linear response density function $\chi(\omega
)$, defined as
\begin{equation}
\chi_{pq,rs}(\omega) = 2 \sum_{I\neq0} \frac{\omega_{I}\left\langle
\Psi_{0}|\hat{a}_{q}^{\dagger}\hat{a}_{p}|\Psi_{I}\right\rangle \left\langle
\Psi_{I}|\hat{a}_{r}^{\dagger}\hat{a}_{s}|\Psi_{0}\right\rangle }{\omega
^{2}-\omega_{I}^{2}}
\end{equation} 
It yields the response of the density matrix to a real one-electron perturbation $\delta\hat{v}$, which in the frequency domain is given  by
\begin{equation}
\delta\gamma_{pq}^{\operatorname{Re}}(\omega)=\sum_{rs}\chi_{pq,rs}
(\omega)\delta v_{rs}(\omega)
\end{equation}
where $\delta\gamma^{\text{Re}}(\omega)$ is the Fourier transform of the real part of the time-dependent response density matrix $\delta\gamma(t)$.

Combining the ERPA formalism for the LR Hamiltonian $\hat{H}^{\rm LR}$ presented in Ref.\citenum{hapka2020long} with the ERPA density response equation derived in Ref.\citenum{drwal2022efficient}, we arrive at the following equation for the ERPA-MC-srDFT response function $\chi^{\rm LR}_{\text{ERPA}}(\omega)$
\begin{equation}
\left[ \omega^{2} - \mathbf{A}_{+}^{\rm LR} \mathcal{N}^{-1} \mathbf{A}_{-}^{\rm LR} \mathcal{N}^{-1} \right] \chi^{\rm LR}_{\text{ERPA}}(\omega) = \mathbf{A}^{\rm LR}_{+} \;
\end{equation}
where, in the representation of the natural spinorbitals corresponding to
$\Psi^{\rm LR}$,
\begin{equation}
\left[ \mathcal{N}\right]_{pq,rs} = \delta_{pr} \delta_{qs} (n_{p}^{\rm LR} - n_{q}^{\rm LR})
\end{equation}
and
\begin{align}
\mathbf{A}_{+}^{\rm LR} & = \mathbf{A}^{\rm LR} + \mathbf{B}^{\rm LR} \\
\mathbf{A}_{-}^{\rm LR} & = \mathbf{A}^{\rm LR} - \mathbf{B}^{\rm LR} + \mathcal{N}
\mathbf{K}\mathcal{N}
\end{align}
The matrices $\mathbf{A}^{\rm LR}$ and  $\mathbf{B}^{\rm LR}$  are defined 
for the Hamiltonian $\hat{H}^{\rm LR}$ and the wavefunction $\Psi^{\rm LR}$ as
\begin{equation}
A_{pqrs}^{\rm LR} = B_{pqsr}^{\rm LR} = \left\langle \Psi^{\rm LR} | [\hat{a}_{p}^{\dag}\hat
{a}_{q},[\hat{H}^{\rm LR},\hat{a}_{s}^{\dag}\hat{a}_{r}]]|\Psi^{\rm LR} \right\rangle
\ \ \ ,
\end{equation}
and are expressible in terms of 1- and 2-RDMs associated with $\Psi^{\rm LR}$, cf.\ Ref.\citenum{hapka2020long}. The matrix $\mathbf{K}$ represents a short-range kernel obtained
from a derivative of the SR potential, cf.\ Eq.~\eqref{LRH}, with respect to the electron density
\begin{equation}
K_{pqrs} = \left\langle \varphi_{p}^{M}\varphi_{q}^{M} | \frac{\delta^{2}
E^{\rm SR}[\rho]}{\delta\rho(\mathbf{r}_{1})\delta\rho(\mathbf{r}_{2})} | \varphi_{r}^{M}\varphi_{s}^{M}\right\rangle
\end{equation}

The ERPA-MC-srDFT linear response  can be viewed as an approximation to the TD-MC-srDFT formalism of Fromager et al.\cite{fromager2013multi}. TD-MC-srDFT reduces to the ERPA-MC-srDFT equations under the assumption that the time-dependent perturbation does not induce relaxation of the configuration-interaction (CI) coefficients, i.e., that only orbital rotations are allowed to respond. For short configuration-expansions of the wavefunction, one expects that both approximations would yield similar results, as confirmed in the next section. 

While TD-MC-srDFT involves coupled orbital and configuration response equations whose dimensionality scales with the size of the CI expansion, ERPA restricts the problem to the orbital-rotation space. As a result, the dimension of the response matrix is dramatically reduced, and the computational cost becomes polynomial in system size rather than combinatorial in active-space size. The computational cost of a direct implementation of ERPA-MC-srDFT scales with the sixth power of the system size.

In practice, TD-MC-srDFT equations are solved using direct iterative techniques\cite{Jorgensen:88,Olsen:88} that avoid the explicit construction and storage of the $\mathbf{A}_{\pm}$ matrices, leading to a cost comparable to second-order MC-srDFT optimization. 
Such direct schemes can be straightforwardly adapted to the ERPA framework. Alternatively, one can employ an iterative algorithm for computing the ERPA response function presented in Ref.~\citenum{daria}, leading to a fifth-power scaling. 
Finally, the computational cost of linear-response MC-srDFT can be further reduced by employing Cholesky decomposition of the two-electron integrals, as recently proposed for TD-MCSCF response by Nottoli, Lipparini, and co-authors.~\cite{Nottoli:25}

\section{Performance of MC-srDFT response for polarizabilities}\label{sec3}

We assess the performance of MC-srDFT in predicting molecular polarizabilities and investigate the impact of tuning the RS parameter on the resulting accuracy. For comparison, we also present results obtained in two limiting cases of the parameter $\mu$: $\mu = 0$, where MC-srDFT reduces to KS-DFT and the linear response becomes equivalent to TD-DFT, and $\mu \rightarrow \infty$, where only the wavefunction component of MC-srDFT remains. If the latter is restricted to a single Slater determinant, the MC-srDFT response is identical to that of time-dependent Hartree-Fock (TD-HF). For genuinely multiconfigurational MC-srDFT, the response function $\chi^{\mathrm{MC\text{-}srDFT}}$ is obtained either from TD-MC-srDFT or from the ERPA-MC-srDFT approaches discussed in the previous section.

We employ a three-component naming convention  to label the results: \textit{Response-WF-srDFA}, where the individual elements of the acronym denote the models used for the linear response, the wavefunction ansatz, and the short-range exchange-correlation density functional approximation (srDFA), respectively. For convenience, all acronyms introduced in this work, together with the corresponding regimes of the range-separation parameter, are summarized in Table~\ref{tab:acro}.

\begin{table*}
\centering
\begin{tabular}{c c c}
\hline
Wavefunction & Short-range XC & Acronym \\ \hline \hline
\multicolumn{3}{c}{WFT limit ($\mu \to \infty$)} \\ \hline
HF     & --     & TD-HF \\
CASSCF & --     & TD-CAS \\
CASSCF & --     & ERPA-CAS \\[4pt]
\hline
\multicolumn{3}{c}{MC-srDFT ($\mu = 0.4$ or opt)} \\ \hline
HF     & srDFA   & TD-HF-srDF \\
CASSCF & srDFA   & TD-CAS-srDF \\
CASSCF & srDFA   & ERPA-CAS-srDF \\[4pt] 
\hline
\multicolumn{3}{c}{DFT limit ($\mu = 0$)} \\ \hline
--     & DFA     & TD-DFT \\[6pt]
\hline
\end{tabular}
\caption{Response models considered in this work together with their acronyms. In cases where both TD-MC-srDFT and ERPA-MC-srDFT equations coincide, the same acronym applies. The short-range exchange-correlation (XC) functional approximation (srDFA) was taken as srLDA\cite{paziani2006local} or srPBE\cite{Goll:05} (see Supporting Information for srPBE results).}
\label{tab:acro}
\end{table*}

The polarizability tensor is computed using the standard expression
\begin{equation}
\alpha_{ij}(\omega) = \sum_{pqrs} \chi^{\text{MC-srDFT}}_{pq,rs}(\omega) \, d^i_{rs} d^j_{pq}\ \ \ 
\end{equation}
where $i,j = \{x,y,z\}$, and $d^i_{rs}, d^j_{pq}$ are dipole-moment matrix elements.

\subsection{Computational Details}\label{sec3a}

The test set comprises ground-state static and dynamic polarizabilities for 14 aromatic molecules reported in Ref.~\citenum{sauer_polari}. Geometries optimized at the MP2/6-31G(d) level for benzene, furan, pyrrole, imidazole, pyridine, pyrimidine, pyrazine, and pyridazine were taken from Ref.~\citenum{Thiel:08}. The geometries of the remaining molecules were obtained at the same level of theory using the Molpro suite of programs.~\cite{Molpro:12}
 
CASSCF reference wavefunctions were chosen as multiconfigurational wavefunctions in MC-srDFT. The MC-srDFT functional, Eq.~\eqref{MCFun}, was optimized self-consistently using the \textsc{Dalton} program.~\cite{dalton} The same software was used to perform TD-CAS, TD-HF, TD-DFT, and TD-CAS-srDFT response calculations. ERPA response calculations were carried out with the \textsc{GammCor} program~\cite{gammcor}, with one- and two-particle reduced density matrices corresponding to the CASSCF reference.

Two types of active spaces were employed. In TD-CAS and ERPA-CAS (the $\mu=\infty$ limit), the active spaces were chosen based on information from the literature~\cite{Thiel:08} and physical insight. For finite RS parameters, we selected active spaces based on MP2-srDFT ($\mu = 0.4\,\mathrm{bohr}^{-1}$) natural occupations\cite{Hubert:16}: all orbitals with occupation numbers $< 1.992$ were considered active. Active spaces chosen this way are typically more compact compared to their $\mu=\infty$ limit,\cite{Kjellgren:19} and thus require fewer computational resources (notice that this is one of the advantages of MC-srDFT compared to the full-range multiconfigurational wavefunction methods). 
For comparison, MC-srDFT results obtained with the physically motivated CASSCF active spaces are also reported and marked with an asterisk (${}^*$). All employed active spaces are listed in Tables~S1 and S2 in the Supplementary Information.

The results presented in the main text were obtained with the LDA\cite{slater1974quantum,LSDA} and srLDA\cite{paziani2006local} functionals. In the Supporting Information, we also report results obtained with the srPBE\cite{Goll:05} model.

The RS parameter $\mu$ was tuned individually for each system. To this end, we performed a discrete scan of $\mu$ in the $0.20$-$0.50$~bohr$^{-1}$ range with a step of $0.05$~bohr$^{-1}$. For each value of $\mu$, the largest (least negative) LR-EKT eigenvalue $\lambda$ was extracted. The optimal $\mu$ was then determined by matching such obtained LR-EKT-based IPs to reference values using one-dimensional piecewise linear interpolation (see Table~S3 in the Supporting Information). Reference IPs were calculated at the PBE0/aug-cc-pVTZ level of theory.

All the polarizabilities were compared with reference CC3\cite{cc3_response} values taken from Ref.~\citenum{sauer_polari}. The frequencies for the dynamic polarizabilities were identical to those used in Ref.~\citenum{sauer_polari}, namely $\omega=$0.072003~a.u. and 0.093215~a.u.

\subsection{Results and Discussion}\label{sec3b}

We begin by investigating the performance of linear-response-based methods which do not employ range separation, i.e.\ the $\mu=0$ and $\mu=\infty$ limiting cases of MC-srDFT. The results are presented in Table~\ref{tab:cas_transposed}.  
TD-CAS, ERPA-CASSCF, and TD-HF systematically underestimate polarizabilities, which can be attributed to the lack of dynamic correlation effects.~\cite{parasuk1996comparison,Champagne_2005} In contrast, TD-LDA overestimates polarizabilities due to the erroneous asymptotic behavior of the LDA exchange-correlation potential.~\cite{Tozer:98,Gruning:01} 
Somewhat surprisingly, TD-CAS performs markedly worse than TD-HF. 
This can be attributed to a cancellation of errors in TD-HF: it overestimates both excitation energies (see, e.g., Ref.~\citenum{hapka2022efficient}) and transition dipole moments 
(cf.\ Figure~S1 in the Supporting Information), leading to an overall reasonable accuracy of the polarizabilities. Replacing TD-HF with TD-CAS, even with relatively small active spaces, disrupts this fortuitous cancellation: the transition dipole moments are reduced, whereas the transition energies (i.e., the denominators in the response function) remain largely unchanged. As a result, the CASSCF polarizability components are smaller in magnitude than the corresponding HF values. 

\begin{table}[ht]
\resizebox{\linewidth}{!}{
\begin{tabular}{l r r r r}
\hline
 & TD-HF & TD-CAS & ERPA-CAS & TD-LDA \\
\multicolumn{5}{l}{static} \\ 
\hline
MAE   &  0.97 &  3.63 &  4.82 & 2.30 \\
ME    & -0.97 & -3.63 & -4.82 & 2.30 \\
SD &  0.49 &  1.08 &  1.17 & 0.81 \\
      &       &       &       &    \\  
\multicolumn{5}{l}{dynamic} \\ \hline
MAE   &  1.09 &  4.15 &  5.81 & 2.73 \\
ME    & -1.09 & -4.15 & -5.81 & 2.73 \\
SD    &  0.57 &  1.64 & 1.52 & 1.03 \\ \hline
\end{tabular}
\caption{Mean error (ME), mean absolute error (MAE), and standard deviation (SD) of the errors in static and dynamic polarizabilities (a.u.) computed with the aug-cc-pVTZ\cite{augDunning} basis set.}
\label{tab:cas_transposed}
}
\end{table}

A comparison of full TD-CAS and ERPA-CAS linear response shows that the former yields slightly more accurate results. The mean absolute errors (MAEs) for the static polarizabilities obtained with TD-CAS and ERPA-CAS amount to 3.6~a.u.\ and 4.8~a.u., respectively. This indicates that the employed active spaces lead to sufficiently large CI expansions of the CASCI wavefunctions such that the response of the expansion coefficients, neglected in ERPA, becomes non-negligible.

The results for dynamic polarizabilities closely resemble those obtained for the static component (Table~\ref{tab:cas_transposed}). Wavefunction-based approaches systematically underestimate the polarizabilities, with TD-HF providing the most accurate results. In contrast, the DFT-based method (TD-LDA) again overestimates the response. Although its accuracy remains inferior to TD-HF, it still outperforms the multiconfigurational methods considered here.

Next, we examine the performance of MC-srDFT with the standard value of the range-separation parameter, $\mu = 0.4\,\mathrm{bohr}^{-1}$ (see Table~\ref{tab:mu04}). 
Increasing $\mu$ from $0$ (the TD-LDA limit) to $0.4\,\mathrm{bohr}^{-1}$ reduces the errors in the polarizabilities for both linear response variants considered.
While TD-LDA systematically overestimates the polarizabilities, CAS-srLDA underestimates them, indicating that $\mu = 0.4\,\mathrm{bohr}^{-1}$ already shifts the results toward the $\mu \to \infty$ limit, where the DFT component vanishes. The TD-CAS-srLDA and ERPA-CAS-srLDA approaches give polarizabilities of practically the same quality. This reflects the fact that compact active spaces in CAS-srLDA lead to short CI expansions of the CAS-srDFT wavefunction. Since static correlation effects are negligible, individual polarizabilities obtained from TD-CAS-srLDA and ERPA-CAS-srLDA calculations deviate from the HF-based values by no more than $0.3~\mathrm{a.u.}$ Note that the inclusion of dynamic correlation in TD-HF-srLDA disrupts the error cancellation observed for TD-HF.

\begin{table*}
\caption{Mean error (ME), mean absolute error (MAE), and standard deviation (SD) of the errors in static and dynamic polarizabilities (a.u.) obtained with range-separated methods at $\mu=0.4\,\mathrm{bohr}^{-1}$ using the aug-cc-pVTZ basis set. CAS$^*$ indicates that a larger active space was employed, as described in Section \ref{sec3a}. 
\label{tab:mu04}} 
\begin{tabular}{l S S S S S} \hline
\multirow{2}{1em}{static}  & \multicolumn{1}{c}{TD-HF-} & \multicolumn{1}{c}{TD-CAS} & \multicolumn{1}{c}{ERPA-CAS} & \multicolumn{1}{c}{TD-CAS*} & \multicolumn{1}{c}{ERPA-CAS*} \\
     &  \multicolumn{1}{c}{-srLDA} & \multicolumn{1}{c}{-srLDA} & \multicolumn{1}{c}{-srLDA} & \multicolumn{1}{c}{-srLDA} & \multicolumn{1}{c}{-srLDA} \\ \hline
ME    &  -1.63 & -1.70 & -1.75 & -1.67 & -1.79 \\
MAE   &   1.63 &  1.70 &  1.75 &  1.67 &  1.79 \\
SD    &   0.54 &  0.53 &  0.56 &  0.57 &  0.55 \\
MAX   &  -1.12 & -1.16 & -1.22 & -1.02 & -1.07 \\
MIN   &  -2.93 & -3.08 & -3.22 & -3.08 & -3.18 \\ 
\multirow{2}{1em}{dynamic}  &  &  &  &  & \\
                            &  &  &  &  & \\ \hline
ME    &  -1.78 & -1.87 & -1.95 & -1.81 & -2.00 \\
MAE   &   1.78 &  1.87 &  1.95 &  1.81 &  2.00 \\
SD    &   0.64 &  0.65 &  0.69 &  0.72 &  0.68 \\
MAX   &  -1.17 & -1.22 & -1.29 & -0.87 & -1.12 \\
MIN   &  -3.58 & -3.87 & -4.12 & -3.87 & -4.03 \\ \hline
\end{tabular}
\end{table*}

Data in Table~\ref{tab:mu04} indicates that there is only a marginal change in the polarizabilities when using the smaller (CAS) active spaces compared with the larger (CAS*) ones. This is expected, since MC-srDFT by construction is less demanding when it comes to the length of the CI expansion than full-range MCSCF methods. Overall, this indicates that not only can the computationally cheaper ERPA approach be used without sacrificing accuracy, but a simpler wavefunction can also be employed, reducing both computational cost and convergence issues.

Even though range separation significantly improves the accuracy for all the CASSCF-based approaches relative to pure CASSCF-based response, the CAS-srDFT errors remain significant. The employment of the srPBE\cite{pbe} functional yields similar results, cf.\ Tables~S4 and S5 in the Supplementary Information. This indicates that the main source of error in the MC-srDFT response is likely the balance between the SR functional and the LR wavefunction; in other words, the chosen value of $\mu$ is not the most favorable.

We applied the optimal tuning of $\mu$ based on the exact condition in Eq.~\eqref{cond}, obtaining $\mu$ values between $0.24\,\mathrm{bohr}^{-1}$ and $0.31\,\mathrm{bohr}^{-1}$ (see Table~S3 in the Supplementary Information). Since the standard choice $\mu = 0.4\,\mathrm{bohr}^{-1}$ underestimates polarizabilities and TD-LDA (the $\mu = 0\,\mathrm{bohr}^{-1}$ limit) overestimates, the optimally-tuned values are expected to improve the MC-srDFT predictions. Indeed, with the use of $\mu_{\text{opt}}$, the errors in all range-separated approaches decrease significantly, with MAE amounting to approximately $0.4\,\mathrm{a.u.}$ for static polarizabilities (see Table~\ref{tab:mu_opt}) and to $0.5\,\mathrm{a.u.}$ for dynamic polarizabilities (see Table~\ref{tab:dyn_mu_opt}). Additionally, the mean errors are close to zero for all three approaches, demonstrating that the optimally tuned $\mu_{opt}$ values effectively eliminate systematic over- or underestimation.

\begin{table*}
\caption{Static ($\omega=0$) polarizabilities (a.u.) compared with the best theoretical estimate (CC3) and experimental values. Mean error (ME), mean absolute error (MAE), and standard deviation (SD) of the errors for range-separated methods with tuned RS parameter $\mu$ obtained using the aug-cc-pVTZ basis set, calculated relative to CC3. Experimental results were taken from $^a$Ref.~\citenum{benzene_exp}, $^b$Ref.~\citenum{furan_exp}, $^c$Ref.~\citenum{azines_exp}, $^d$Ref.~\citenum{pyrrole_exp}, respectively.\label{tab:mu_opt}}
\begin{tabular}{l SSSSS}
\hline
 &
  \multicolumn{1}{l}{TD-HF-srLDA} &
  \multicolumn{1}{l}{TD-CAS-srLDA} &
  \multicolumn{1}{l}{ERPA-CAS-srLDA} &
  \multicolumn{1}{l}{CC3} &
  \multicolumn{1}{l}{Experiment} \\ \hline
benzene      & 68.28 & 68.40 & 68.40 & 68.49 & 67.48$^a$                \\
benzonitrile & 87.26 & 87.47 & 87.46 & 85.66 &                      \\
furan        & 48.42 & 48.47 & 48.45 & 48.34 & 48.59$^b$                \\
imidazole    & 48.61 & 48.68 & 48.66 & 49.17 &                      \\
oxazole      & 43.13 & 43.18 & 43.17 & 43.18 &                      \\
phenol       & 74.71 & 74.83 & 74.82 & 74.15 &                      \\
pyrazine     & 58.67 & 58.52 & 58.52 & 58.83 & 60.62$^c$                  \\
pyridazine   & 58.51 & 58.45 & 58.43 & 58.73 & 59.32$^c$                  \\
pyridine     & 63.24 & 63.23 & 63.23 & 63.19 & 64.11$^c$                  \\
pyrimidine   & 57.82 & 57.78 & 57.78 & 57.81 & 59.35$^c$                \\
pyrrole      & 54.05 & 54.11 & 54.09 & 54.47 & 53.47$^d$                 \\
phosphole    & 72.15 & 72.44 & 72.45 & 73.52 & \multicolumn{1}{l}{} \\
thiazole     & 58.35 & 58.62 & 58.55 & 58.87 & \multicolumn{1}{l}{} \\
thiophene    & 63.56 & 63.69 & 63.69 & 63.85 & 65.18$^b$                 \\ 
\hline
\bf{ME }          & -0.11 & -0.03 & -0.04 &       & \multicolumn{1}{l}{} \\ 
\bf{MAE }         & 0.44  & 0.41  & 0.42  &       & \multicolumn{1}{l}{} \\ 
\bf{SD}     & 0.66  & 0.65  & 0.66  &       & \multicolumn{1}{l}{} \\ \hline
\end{tabular}
\end{table*}

\begin{table*}
\begin{tabular}{l SSSS}
\hline
 & \multicolumn{1}{c}{TD-HF-srLDA} 
 & \multicolumn{1}{c}{TD-CAS-srLDA}
 & \multicolumn{1}{c}{ERPA-CAS-srLDA}
 & \multicolumn{1}{c}{CC3} \\ \hline
$\mathbf{\omega=0.072003\,\mathrm{a.u.}}$     &          &           &            &       \\ 
benzene        & 70.71    & 70.85     & 70.85      & 70.86 \\
benzonitrile   & 90.90    & 91.14     & 91.13      & 88.94 \\
furan          & 49.88    & 49.93     & 49.91      & 49.75 \\
imidazole      & 50.04    & 50.12     & 50.09      & 50.65 \\
oxazole        & 44.30    & 44.36     & 44.33      & 44.32 \\
phenol         & 77.50    & 77.63     & 77.62      & 76.82 \\
pyrazine       & 60.75    & 60.57     & 60.57      & 60.89 \\
pyridazine     & 60.44    & 60.38     & 60.37      & 60.62 \\
pyridine       & 65.40    & 65.39     & 65.38      & 65.31 \\
pyrimidine     & 59.66    & 59.61     & 59.61      & 59.62 \\
pyrrole        & 55.81    & 55.88     & 55.86      & 56.25 \\
phosphole      & 75.28    & 75.59     & 75.58      & 76.85 \\
thiazole       & 60.17    & 60.49     & 60.38      & 60.69 \\
thiophene      & 65.71    & 65.84     & 65.84      & 65.98 \\
\pagebreak
$\mathbf{\omega=0.093215\,\mathrm{a.u.}}$     &          &           &            &       \\ 
benzene        & 72.54    & 72.69     & 72.69      & 72.63 \\ 
benzonitrile   & 93.72    & 93.98     & 93.97      & 91.45 \\
furan          & 50.97    & 51.02     & 51.00      & 50.80 \\
imidazole      & 51.10    & 51.19     & 51.16      & 51.76 \\
oxazole        & 45.16    & 45.23     & 45.20      & 45.16 \\
phenol         & 79.63    & 79.77     & 79.77      & 78.85 \\
pyrazine       & 62.38    & 62.17     & 62.16      & 62.50 \\
pyridazine     & 61.94    & 61.89     & 61.88      & 62.07 \\
pyridine       & 67.02    & 67.01     & 67.01      & 66.91 \\
pyrimidine     & 61.04    & 60.99     & 60.99      & 60.98 \\
pyrrole        & 57.13    & 57.20     & 57.18      & 57.59 \\
phosphole      & 77.76    & 78.07     & 78.04      & 79.51 \\
thiazole       & 61.53    & 61.90     & 61.74      & 62.05 \\
thiophene      & 67.33    & 67.47     & 67.46      & 67.58 \\ \hline
\textbf{ME}  & -0.06    & 0.03      & 0.01       &       \\
\textbf{MAE}   & 0.51     & 0.48      & 0.49       &       \\
\textbf{SD} & 0.81     & 0.81      & 0.82       &       \\ \hline
\end{tabular}
\caption{Dynamic polarizabilities (a.u.) compared to the best theoretical estimate (CC3). Mean error (ME), mean absolute error (MAE), and standard deviation (STDEV) of the errors for range-separated methods with tuned RS parameter $\mu$ in the aug-cc-pVTZ basis set, calculated relative to CC3. \label{tab:dyn_mu_opt}}
\end{table*}

\begin{figure}
    \centering
    \includegraphics[width=0.99\linewidth]{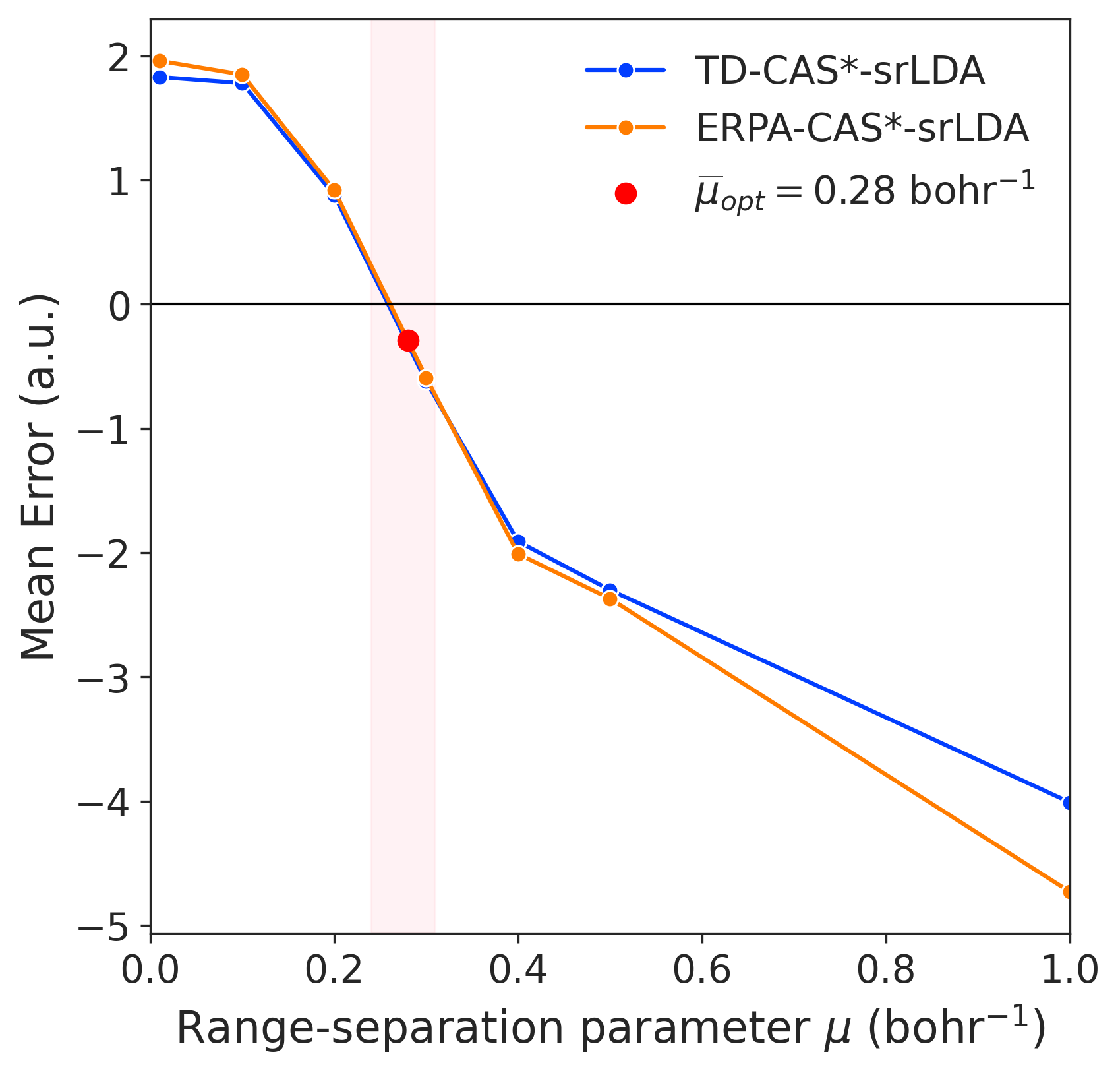}
    \caption{Mean error of static polarizabilities calculated in the aug-cc-pVDZ basis set. The area highlighted in red marks the range of tuned $\mu$ values for the studied systems, while the red dot marks the average tuned value $\bar{\mu}_{opt}=0.28\,\mathrm{bohr}^{-1}$. Pirydazine was excluded from the set due to poor convergence of the CASSCF wave function in this basis set. CAS$^*$ denotes that the larger active space was used, as described in Section \ref{sec3a}. }
    \label{fig:scan}
\end{figure}

In Figure~\ref{fig:scan}, we show the mean error as a function of the RS parameter. The minimum  lies within the interval of the optimally tuned $\mu$ values. The average  tuned parameter, $\bar{\mu}_{\text{opt}}=0.28\,\mathrm{bohr}^{-1}$, is close to the value that minimizes the mean error (ca.\ $0.25\,\mathrm{bohr}^{-1}$). Thus, we recommend using $\mu=0.28\,\mathrm{bohr}^{-1}$ for ground-state polarizability calculations as a practical alternative to the more computationally demanding EKT-based tuning procedure.

\begin{figure}
    \centering
    \includegraphics[width=0.99\linewidth]{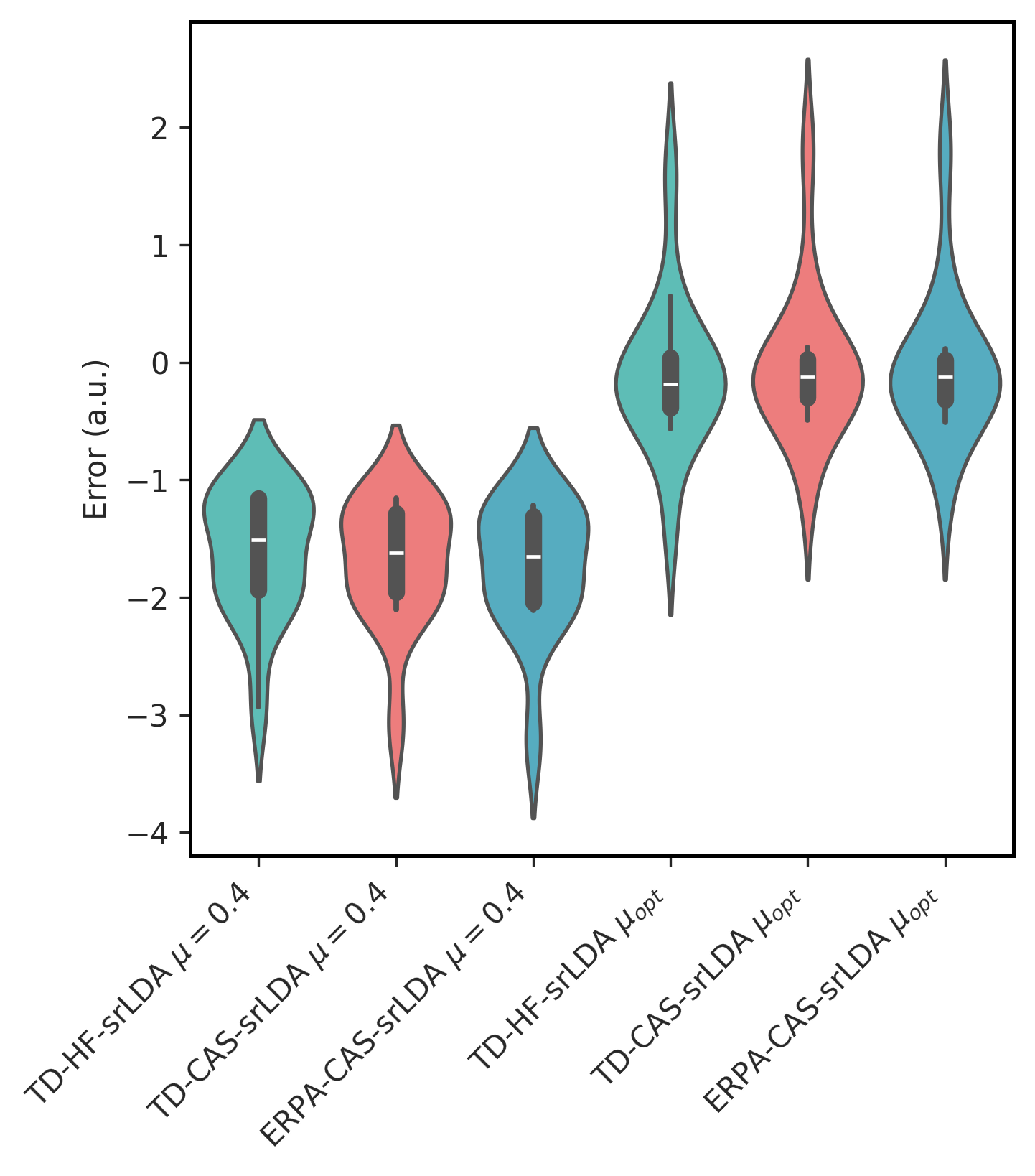}
    \caption{Violin plots of errors in static polarizability for range-separated methods with a standard value of the range-separation parameter ($\mu=0.4\,\mathrm{bohr}^{-1}$) vs. the optimally tuned values, $\mu_{opt}$.}

    \label{fig:violin}
\end{figure}

Figure~\ref{fig:violin} shows the distribution of errors in a violin plot. For $\mu_{\rm opt}$, unlike for $\mu=0.4\,\mathrm{bohr}^{-1}$, the distributions of error for all three methods are approximately symmetric and unimodal, and the mean values are close to zero. The two outliers are benzonitrile and phosphole, the first having an error of $1.8~\mathrm{a.u.}$ for both TD-CAS-srLDA and ERPA-CAS-srLDA, and the second $-1.1\, \mathrm{a.u.}$ with optimally tuned values of $\mu_{\rm opt}=0.240\,\mathrm{bohr}^{-1}$ and $\mu_{\rm opt}=0.263\,\mathrm{bohr}^{-1}$, respectively. These errors are still within an acceptable range.

\section{Summary and Conclusions}\label{sec:summary}

We have proposed a range-separation parameter tuning procedure tailored to multiconfiguration short-range DFT. By generalizing the derivation of Morrell, Parr, and Levy,\cite{mpl} we have shown that the $\mu$ value can be determined to enforce the correct exponential decay of the approximate MC-srDFT electron density with the first ionization potential of a given system. Importantly, this ionization potential emerges as the largest (least negative) eigenvalue of the Extended Koopmans' Theorem\cite{morrison1992extended,smith1975extension} matrix constructed from the long-range Hamiltonian and the MC-srDFT reduced density matrices.

We have analyzed the performance of optimally-tuned MC-srDFT employing CASSCF wave functions for static and dynamic polarizabilities of ground-state molecular systems taken from Ref.\citenum{sauer_polari}. Two variants were considered: the full linear-response formalism (TD-MC-srDFT)\cite{fromager2013multi} and its approximate counterpart based on the extended random phase approximation (ERPA-srDFT). Optimal tuning of CAS-srLDA leads to a clear improvement over the commonly used universal value $\mu=0.4$~bohr$^{-1}$.~\cite{Fromager:07} The mean absolute TD-CAS-srLDA error with respect to the CC3 benchmark is reduced from 1.7 to 0.4~a.u.\ and from 1.9 to 0.5~a.u.\ for the static and dynamic polarizabilities, respectively. The srLDA and srPBE functionals exhibit similar performance, in agreement with previous observations for excitation energies\cite{Kjellgren:19} and indirect spin-spin couplings.~\cite{Kjellgren:21,Hapka:25} Since the optimally tuned $\mu$ values fall within a relatively narrow range, we recommend $\mu=0.28$~bohr$^{-1}$ for ground-state polarizabilities as a practical alternative to system-specific tuning. 

ERPA-srDFT gives results nearly identical to those obtained with the full TD-CAS-srDFT linear response. This stands in contrast to the comparison between the methods in the $\mu \rightarrow \infty$ limit, i.e.,\ ERPA-CAS and TD-CAS, where ERPA is noticeably less accurate. The difference is readily understood: the inclusion of dynamical correlation via srDF leads to more compact CI expansions, so that response contributions from the expansion coefficients become less important. Irrespective of the linear-response framework employed, MC-srDFT with the optimally tuned RS parameter greatly improves upon both of its $\mu$ limits, namely the systematic underestimation of polarizabilities by pure wavefunction approaches (TD-CAS and TD-HF) and the overestimation characteristic of TD-DFT.
 
Although the present study focuses on electronic ground states, the proposed EKT-based optimal-tuning MC-srDFT framework is readily extendable to excited states. 
Investigations in this direction are currently underway.


\begin{acknowledgement}
The National Science Center of Poland supported this work under grants no.\ 2021/43/I/ST4/02250 and no.\ 2021/43/D/ST4/02762. For the purpose of Open Access, the authors have applied a CC-BY public copyright license to any Author Accepted Manuscript (AAM) version arising from this submission.
\end{acknowledgement}

\vspace{0.5cm}

\bibliography{biblio}

\end{document}